# Analytical reasoning task reveals limits of social learning in networks

IYAD RAHWAN, Masdar Institute of Science & Technology, UAE; University of Edinburgh, UK
DMYTRO KRASNOSHTAN, Masdar Institute of Science & Technology, UAE
AZIM SHARIFF, University of Oregon, USA
JEAN-FRANÇOIS BONNEFON, CNRS & Université de Toulouse, France

1. INTRODUCTION

Social learning—by observing and copying others—is a highly successful cultural mechanism for adaptation, outperforming individual information acquisition and experience [Rendell et al. 2010]. This ability is important for the spreading of best practices [Pentland 2012], useful information [Mason and Watts 2012], healthy habits [Centola 2010], and cooperation [Fowler and Christakis 2010].

Here, we investigate social learning in the context of the uniquely human capacity for reflective, analytical reasoning. A hallmark of the human mind is our ability to engage analytical reasoning, and suppress false associative intuitions [Kahneman 2011]. Networks can serve two purposes in relation to analytic reasoning. First, networks may *propagate analytical reasoning processes*. That is, individuals who witness rational decisions going against their intuition may be prompted to reflect and spontaneously switch to a more analytic thinking style in subsequent, similar tasks. We refer to this phenomenon as the *contagion of analytical processing*. Another possibility is that networks *propagate correct responses to analytic problems*. That is, individuals who witness rational decisions going against their intuition may recognize their intuition as incorrect, and adopt the correct decision, but do so without engaging analytic reasoning themselves. Thus, increased connectivity, by increasing the availability of diverse information sources, may enable individuals to obtain higher-quality information and perspectives, without necessarily being able to generate similar insights independently. We refer to this phenomenon as the *contagion of analytical output*.

We ran 5 lab-based network sessions, involving 20 subjects each. In each session, subjects sat at individual computer workstations and solved a series of analytic problems. Each subject was randomly assigned to a node in an underlying network, which determined the neighbors (in the sense of the network, rather than physical proximity) whose responses were visible to the subject. Different network topologies were used in five sessions (20 participants each). The first session provided a baseline condition in which subjects were not connected to any neighbor, and thus did not see any of the other participants' responses. The other sessions spanned a wide range of possible structures: fully connected, Erdős-Rényi random graph, Barabási-Albert graph with hubs, and a highly clustered graphs.

Subjects were asked to solve a series of 3 questions (see table) known as the Cognitive Reflection Test (CRT) [Frederick 2005]. They require engaging analytic reasoning to overcome an incorrect intuition. No particular skill or knowledge is required to generate the correct answer – only the engagement of effortful, analytic reasoning process. Thus, there is no particular 'trick' which, once learned, can be used in subsequent tasks. The subject should simply recognize that initial intuition cannot be trusted, and a more reflective attitude is needed.

| Question | Intuitive | Correct |
| --- | --- | --- |
| In a lake, there is a patch of lily pads. Every day, the patch doubles in size. If it takes 48 days for the patch to cover the entire lake, how long would it take for the patch to cover half of the lake? | 24 | 47 |
| If it takes 5 machines 5 minutes to make 5 widgets, how many minutes would it take 100 machines to make 100 widgets? | 100 | 5 |
| A bat & a ball cost $1.10 in total. The bat costs $1.00 more than the ball. How much does the ball cost? | 10 c | 5 c |

Each subject answered 5 trials for each of the 3 questions. In the first trial, subjects responded independently. In the subsequent trials 2 to 5, subjects could see the responses that their network neighbors entered in previous rounds. No information was given about the accuracy of these responses. Subjects were informed that they would accumulate monetary rewards for every correct response they gave, on every trial. This setup provides an ideal test-bed to pit analytical process contagion against analytical output contagion. Output contagion should improve performance from one trial to the next (within each question), but not from one question to the next. Process contagion should improve performance from one question to the next, in addition to improving performance from one trial to the next.

## 2. RESULTS

Subjects' performance appears in the figure below. Separate logistic regressions were conducted in each topology. To detect process contagion, we tested whether the performance of subjects in each of our four topologies improved across questions, over and above the progression observed in the baseline condition. For example, in the case of the Clustered topology, we conducted a logistic regression in which the predictors were the question (first, second, third), the topology (Baseline, Clustered), and their interaction. The dependent measure was the performance (correct or incorrect) during the first trial of each question. What counts as evidence for process contagion is a significant interaction between question and topology, showing that increase in performance in the network group is greater than increase in performance in the Baseline group. We detected no such significant interaction for any topology, all $z < 1.05$, all $p > .28$. Performance never improves significantly from one question to the next.

To detect output contagion, we tested whether the performance of subjects in each of our four topologies improved across trials within each question, above the progression observed in the Baseline. For example, in the case of the Clustered topology, we conducted a logistic regression in which the predictors were the trial (first, last), the topology (Baseline, Clustered), and their interaction. What counts as evidence for process contagion is a significant interaction between trial and topology, showing that increase in performance in the network group is greater than in the Baseline group. We obtained such evidence for all topologies except Clustered. In all other topologies, subjects' performance largely improved across trials, as the correct response to each question spread in turn across the network.

The Clustered topology was an exception insofar as it seemed unable to improve performance over the Baseline group. One possible reason might be that connectivity in the Clustered network is insufficient to spread the correct, analytical response. To test whether the individual connectivity of a node was linked to the final performance of the subject in this node, we computed an index of connectivity (global distance to all other nodes, i.e., closeness centrality) and an index of final performance (average proportion of correct responses during the last trial of each question), for each node in each network. As expected, they were significantly correlated, $r(78) = .38$, $p < .0001$.

## 3. DISCUSSION

Our data show that networks can help to solve analytic problems – but with a caveat. Networks do not propagate the analytic reasoning style required to independently arrive at correct answers. They can only propagate the correct response to analytic problems, one at a time. This failure to propagate analytical processing is striking. Consider that it is possible to prime analytical processing using very subtle cues—such as an evocative image of Rodin's Thinker [Gervais and Norenzayan 2012] or listing questions using a challenging font [Alter et al. 2007]. How can we explain, then, that repeated exposure to the analytic output of peers in a network, and even the subsequent recognition and adoption of their correct answer, all fail to prime analytic reasoning in subsequent tasks?

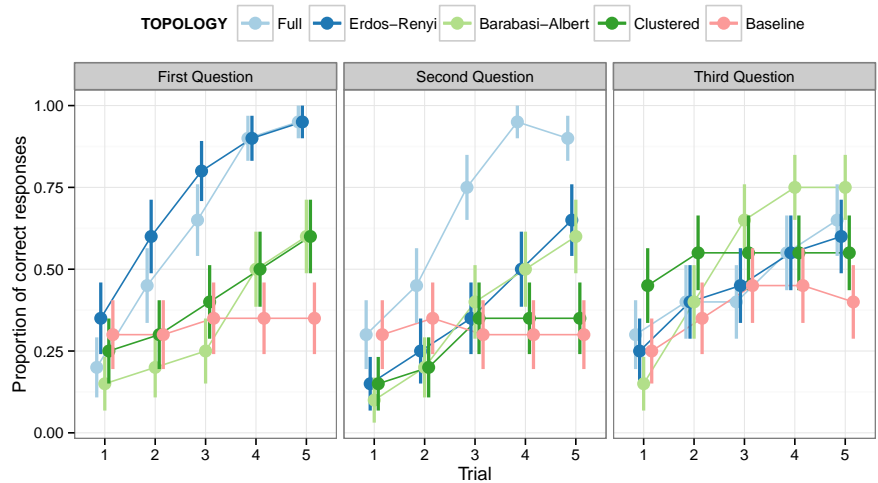

Social learning is a low-cost phenomenon because learners evaluate behaviors without necessarily understanding what makes a behavior successful. The trade-off, though, is that without that deep understanding, learners can be inaccurate in what they choose to copy [Boyd et al. 2011]. This propensity may explain why subjects persist in copying only analytical responses in our tasks, whilst copying analytical processing would be fairly easy, cost-less, and financially rewarding. We have therefore identified an *unreflective copying bias*, thus contributing to our understanding of the limits of increased connectivity in amplifying collective intelligence [Sparrow et al. 2011; Lorenz et al. 2011].